\documentclass[conference]{IEEEtran}
\IEEEoverridecommandlockouts
\usepackage{cite}
\usepackage{amsmath,amssymb,amsfonts}
\usepackage{algorithmic}
\usepackage{graphicx}
\usepackage{textcomp}
\usepackage{xcolor}
\def\BibTeX{{\rm B\kern-.05em{\sc i\kern-.025em b}\kern-.08em
    T\kern-.1667em\lower.7ex\hbox{E}\kern-.125emX}}
\begin{document}

\title{A 4-Element 800MHz-BW 29mW True-Time-Delay Spatial Signal Processor Enabling Fast Beam-Training with Data Communications  \\
%{\footnotesize \texsuperscript{*}Note: Sub-titles are not captured in Xplore and should not be used}
\thanks{Supported partly by NSF awards (\#: 1718742, 1955672, 955306, and 1944688), CDADIC, DARPA JUMP ComSenTer and CONIX centers.  }
}
\author{
\IEEEauthorblockN{Chung-Ching Lin, Chase Puglisi, Soumen Mohapatra, \\ Erfan Ghaderi, Deukhyoun Heo, Subhanshu Gupta}
\IEEEauthorblockA{\textit{School of Electrical Engineering and Computer Science} \\
\textit{Washington State University, Pullman, USA}\\
chung-ching.lin@wsu.edu}
\and
\IEEEauthorblockN{Veljko Boljanovic, Han Yan, Danijela Cabric}
\IEEEauthorblockA{\textit{Electrical and Computer Engineering Department} \\
\textit{University of California Los Angeles}\\
Los Angeles, USA \\
vboljanovic@ucla.edu }
}

\maketitle

\begin{abstract}
Spatial signal processors (SSP) for emerging millimeter-wave wireless networks are critically dependent on link discovery. To avoid loss in communication, mobile devices need to locate narrow directional beams with millisecond latency. In this work, we demonstrate a true-time-delay (TTD) array with digitally reconfigurable delay elements enabling both fast beam-training at the receiver with wideband  data communications. In beam-training mode, large delay-bandwidth products are implemented to accelerate beam training using frequency-dependent probing beams. In data communications mode, precise beam alignment is achieved to mitigate spatial effects during beam-forming for wideband signals. The 4-element switched-capacitor based time-interleaved array uses a compact closed-loop integrator for signal combining with the delay compensation implemented in the clock domain to achieve high precision and large delay range. Prototyped in TSMC 65nm CMOS, the TTD SSP successfully demonstrates unique frequency-to-angle mapping with 3.8ns maximum delay and 800MHz bandwidth in the beam-training mode. In the data communications mode, a $\sim12$dB uniform beamforming gain is achieved from 80MHz to 800MHz. The TTD SSP consumes 29mW at 1V supply achieving 122MB/s with 16-QAM at $9.8\%$ EVM. 
\end{abstract}

\begin{IEEEkeywords}
spatial signal processor, true-time-delay, frequency-to-angle mapping, ring amplifiers, fast beam training
\end{IEEEkeywords}

\section{Introduction}
%Large antenna arrays for space, weather, and infrastructure applications are conventionally implemented using analog/hybrid antenna arrays driven by the need for lower costs and power consumption \cite{fikes2020}. Despite clear implementation advantages, these arrays are challenged by a combination of system-design parameters including the number of antennas, modulation bandwidth, steering angles, and tracking latency \cite{chu2013, ghaderi2019}. These parameters further impact critical signal processing functions such as beam-training and beam-forming that are of paramount importance in applications on-the-move demanding high bandwidth with low latency. 

\begin{figure}[t]
\begin{minipage}[t]{0.47\linewidth}
    \centering
    \includegraphics[width=0.89\textwidth]{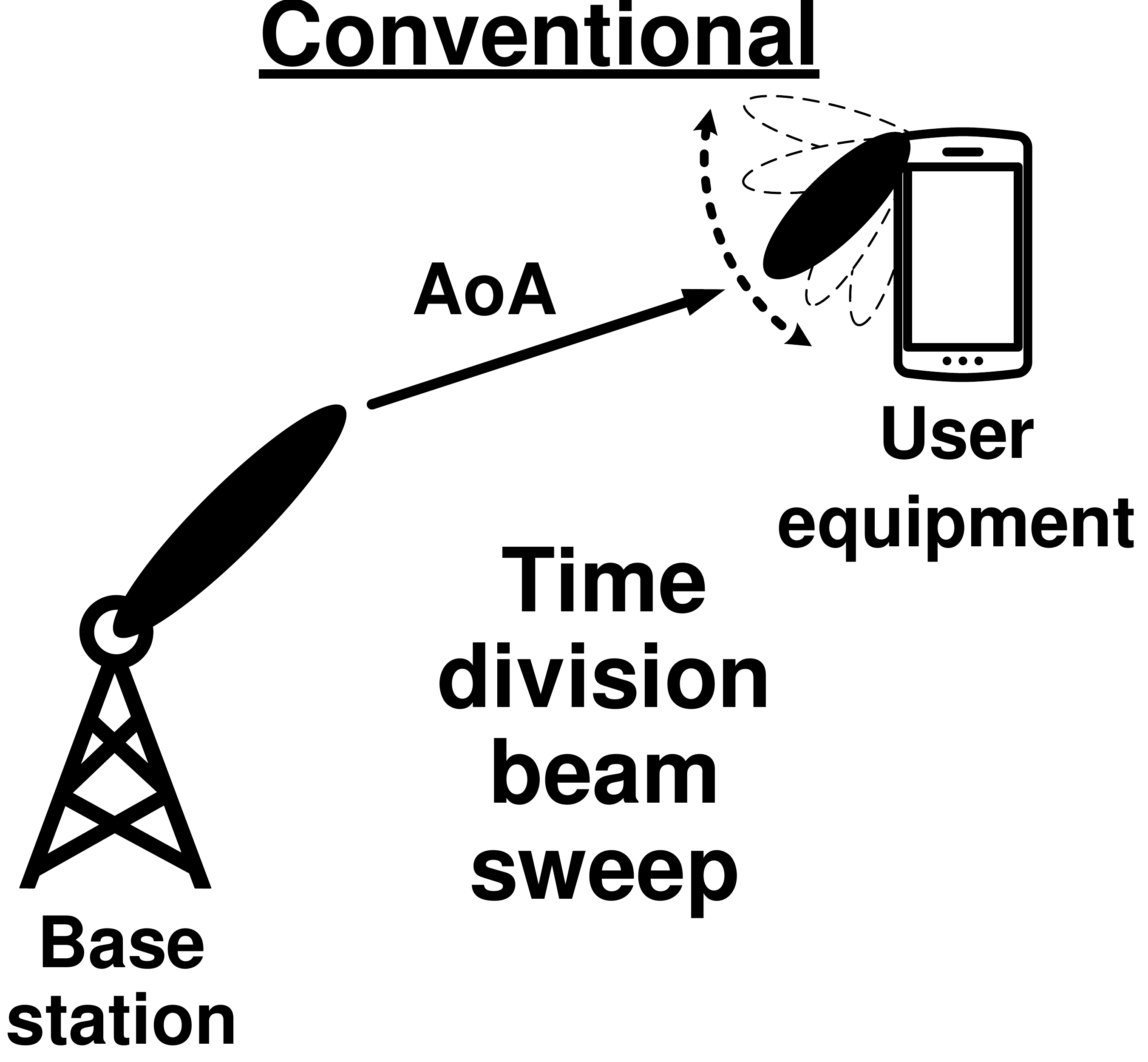}
    (a)
\end{minipage}
\hspace{0.1cm}
\begin{minipage}[t]{0.47\linewidth} 
    \centering
    \includegraphics[width=0.89\textwidth]{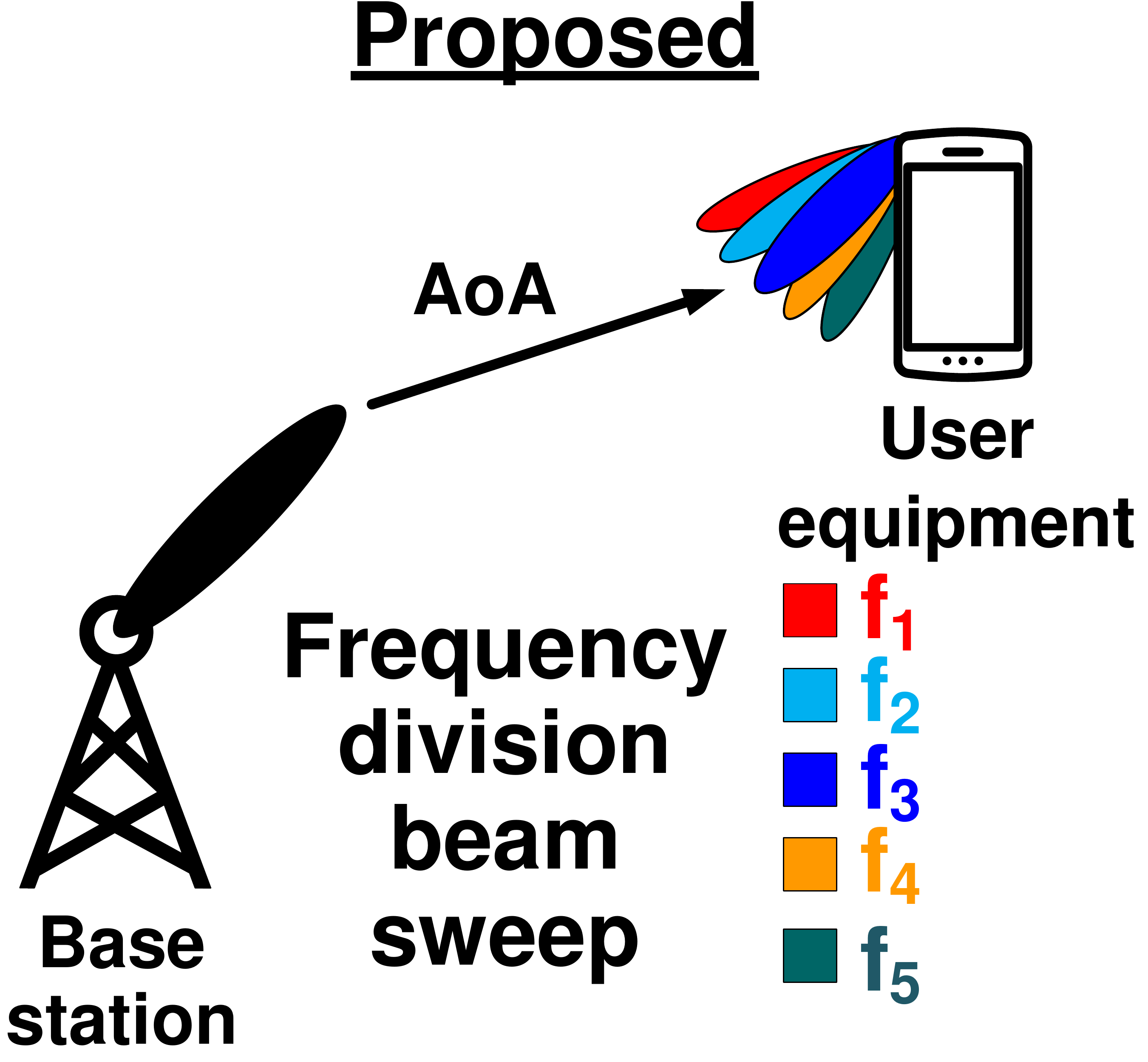}
    (b)
\end{minipage}
\caption{(a) Latency-costly time-division beam sweep in analog arrays , and (b) proposed low-latency frequency-division beam sweep in analog arrays.}
\vspace{-7mm}
\label{fig:PATDD}
\end{figure}

Wireless communication and sensing applications at millimeter-wave (mmWave) frequencies require rapid localization and direction finding of mobile nodes \cite{ghasempour_2020, Boljanovic2021}.  However, current protocols for direction finding are based on repetitive time-division beam sweeping techniques that are often non-scalable and costly in latency (Fig.~\ref{fig:PATDD}(a)). Novel array architectures and signal processing techniques are required to avoid prohibitive beam training overhead associated with large antenna arrays and narrow beams at mmWave.

%Emerging wireless networks at millimeter-wave require large antenna arrays \cite{fikes2020} to compensate for propagation losses. Analog/hybrid phased arrays while providing lower implementation costs and energy efficiencies are limited by tracking latency and fractional bandwidths. This limitation arises from approximating the inter-element time delays with  phase shift. Specifically, beam training (direction-finding), at the network edge nodes, needs to be performed through exhaustive latency-costly beam sweeping, while directional data communication suffers from the wideband spatial effect (beam squint). These limitations worsens linearly/quadratically in 1D 2D  arrays  \cite{Yan:TTD, Boljanovic2021} posing serious challenges for low-latency applications requring high mobility and wide bandwidths. Fig.~\ref{fig:PATDD}(a) highlights the conventional beam sweeping method that makes it non-scalable in large arrays. 

True-time-delay (TTD) arrays are appealing yet insufficiently investigated alternative for both fast mmWave beam training  and wideband directional data communications. Due to time delaying of the signal in each antenna branch, TTD arrays have frequency-dependent probing beams, which can be exploited to enhance the channel probing capability or  mitigate spatial effects by fully adjusting the delay introduced in the TTD circuits. Early implementations relied on delay lines in all antenna branches \cite{chu2013}, but this approach suffers from limited delay ranges when implemented at RF which renders it insufficient to achieve frequency dispersive beam training as proposed in this work. Recent advancement in TTD arrays with baseband delay elements and large delay range-to-resolution ratios \cite{ghaderi2019, Ghaderi2020ISSCC} improved the scalability and thus enabled the realization of fast beam training schemes in large arrays without needing any bulky and costly external constructs (such as Rotman Lens \cite{Rotman2016} or Leaky Wave Antenna \cite{Saeidi2021}). Lastly, it does not have wideband spatial effects (beam-squint) inherent to wideband phase-shift based arrays.

This work presents a reconfigurable TTD spatial signal processor (SSP) with large delay-bandwidth product for unique frequency-to-angle mapping for fast beam-training. Additionally, Wideband data communications is demonstrated using large unity-gain bandwidth and fast slewing ring amplifiers. The proposed architecture validates our fast beam-training algorithm for analog / hybrid arrays proposed in \cite{Yan:TTD, Boljanovic2021} with hardware implementation for the first time. The salient features of this work in the two operation modes are as follows. In the beam-training mode, we demonstrate unique frequency-to-angle mapping with wide field-of-view for fast frequency division based beam-training. The proposed beam-training method do not need costly/bulky external constructs such as Rotman Lens \cite{Rotman2016} or Leaky-Wave Antenna \cite{Saeidi2021}. In the data communication mode, we achieve a 60\% increase in modulation bandwidth compared to prior art consuming 33\% lower power. The TTD SSP uses a 5mW 800MHz bandwidth ring amplifier  based switched-capacitor integrator with delay compensation to achieve large delay-bandwidth product. 

% The rest of the paper is as follows. Section II describe the TTD-based beam training and data communications algorithm. Section III presents the proposed SSP architecture and circuit design techniques for the two modes. Section IV presents the measured results followed by conclusions in Section V.

% \begin{figure}[t]
% \begin{minipage}[t]{0.48\linewidth}
%     \centering
%     \includegraphics[width=1\textwidth]{Fig_ESSCIRC/beam_training.pdf}
%     (a)
% \end{minipage}
% \hspace{0.1cm}
% \begin{minipage}[t]{0.48\linewidth} 
%     \centering
%     \includegraphics[width=1\textwidth]{Fig_ESSCIRC/beamforming.pdf}
%     (b) 
% \end{minipage}
% \caption{Beams for frequencies $f_1,...,f_4$, designed for (a) beam training and (b) directional data communication with a 4-element TTD array.}
% \label{fig:beams}
% \end{figure}  

%\begin{figure}[t]
%\begin{minipage}[t]{0.48\linewidth}
%    \centering
%    \includegraphics[width=1\textwidth]{Fig_ESSCIRC/BT_Tau.pdf}
%    (a)
%\end{minipage}
%\hspace{0.1cm}
%\begin{minipage}[t]{0.48\linewidth} 
%    \centering
%    \includegraphics[width=1\textwidth]{Fig_ESSCIRC/BF_Tau.pdf}
%    (b) 
%\end{minipage}
%\caption{Delay settings for (a) beam training and (b) directional data %communication with a 4-element TTD array.}
%\label{fig:tau}
%\end{figure}  

\section{TTD-based Low-Latency Beam Training with Wideband Communications}

% Short intro
%In our recent work, we proposed the use of TTD arrays to address the problems of mmWave beam training \cite{Yan:TTD, Boljanovic2020, Boljanovic2021} and directional beam steering in data communication \cite{ghaderi2019,Ghaderi2020ISSCC}.  
This section presents the system design for both wideband beam training and data communication, and the associated challenges in TTD IC design. We focus on uniform linear array with $N$ elements. The channel dominant propagation direction is denoted as $\theta$ and the array response is $[\mathbf{a}(\theta,f)]_n = \exp(-j2\pi(n-1) d f \sin(\theta)/c)$, where $d$ is the element spacing, $f$ is the frequency, and $c$ is the speed of light.

% Beam training
\textbf{Low-latency beam training:} The objective of beam training is to use signal received from search beams to identify $\theta$. Fig.~\ref{fig:PATDD}(a) illustrates the conventional beam training for phased array, where one direction is searched at a time. Hence, finding $\theta$ involves high search latency which becomes more troublesome with larger array size and narrower beam-width in the future. Fig.~\ref{fig:PATDD}(b) shows our recently proposed fast beam training tailored for TTD array. Our approach leverages frequency-dependent search beam to sound all directions simultaneously which
% using a single wideband pilot in a multi-carrier system 
greatly reduces beam training latency \cite{Boljanovic2021}. More specifically, when TTD delay taps are set to be $\tau_n$ in the $n$-th element of a uniform linear array, the search beam synthesized at the frequency $f$ is defined by the antenna weight vector $\mathbf{w}(f)$, whose $n$-th element is $[\mathbf{w}(f)]_n=\exp(-j2\pi f \tau_n)$. Different incident angle $\theta$ leads to different system frequency response $|\mathbf{w}^{\text{H}}(f)\mathbf{a}(\theta,f)|^2$. Hence, one can infer unknown $\theta$ by first measuring the system's frequency response, which can be achieved by a single multi-carrier pilot symbol. Furthermore, with the signal bandwidth $\text{BW}$, our theory shows that setting the TTD delay taps as $\tau_n=(n-1)/\text{BW},~n=1,...,N$ enables searching the entire angular range $\theta \in [-\pi/2,\pi/2]$ at once \cite{Boljanovic2021}.
% Upon receiving a wideband pilot, the information of the optimal angle/beam is encoded in the spectrum of received signal and it can be acquired by identifying the frequency component with the strongest received power due the to one-to-one frequency-to-angle mapping. With a wide mmW bandwidth in the order of $\text{BW}=1$GHz and delay difference between neighboring elements in the order of $\Delta\tau = 1/\text{BW} = 1$ns, the proposed beam training have a relaxed delay resolution requirement. The proposed beam training can be easily accommodated with the state-of-the-art TTD hardware, which support resolutions in the order of several picoseconds \cite{ghaderi2019}. 
% Although appealing in reducing the beam training latency, 
The approach required the maximum TTD delay to be $\tau_{N}=(N-1)/\text{BW}$. For example,  the delay range needs to be $3.75$ns given $800$MHz signal bandwidth and the array with $N=4$ elements. Such bandwidth and delay range requirement represents a big design challenge of TTD IC.
% Specifically, given the maximum delay compensation $T_{\text{C-max}}$, which is in the order of $\sim15$ns in the state-of-the-art TTD hardware, the training can be realized if the condition $\tau_N=(N-1)/\text{BW}\leq T_{\text{C-max}}$ is satisfied. 
% Thus, with fixed $T_{\text{C-max}}$, the design and performance of TTD-based beam training depends on fundamental trade-offs between the system parameters $N$ and $\text{BW}$ \cite{Boljanovic2021}.

% Data communication

\textbf{Wideband communication:} After the propagation direction $\theta$ is identified, TTD-based directional data communication aligns all the signal frequency components with $\theta$ and thus avoid the beam squint effect in wideband mmWave systems. Mathematically, the delay taps $\tau_n$ needs to be designed such that antenna weight $\mathbf{w}(f)$ matches with array spatial response $\mathbf{a}(\theta,f)$ for all frequencies, which requires the delay taps $\tau_n = (n-1)d\sin(\theta)/c$. In contrast to beam training, this task is mainly constrained by achievable delay resolution.
% , which often needs to be in the order of several picoseconds at mmWave frequencies.

%\textcolor{blue}{VB: Maybe we can include a short paragraph here and highlight how our current testbed can facilitate both operational modes (beam training and data communication), because the chip has a good delay-range ratio. Also, the text that comes before can be shortened and/or modified. We can include some more hardware-oriented comments if you prefer that.}

\begin{figure}[t]
\centering
         
    \includegraphics[width=0.9\columnwidth]{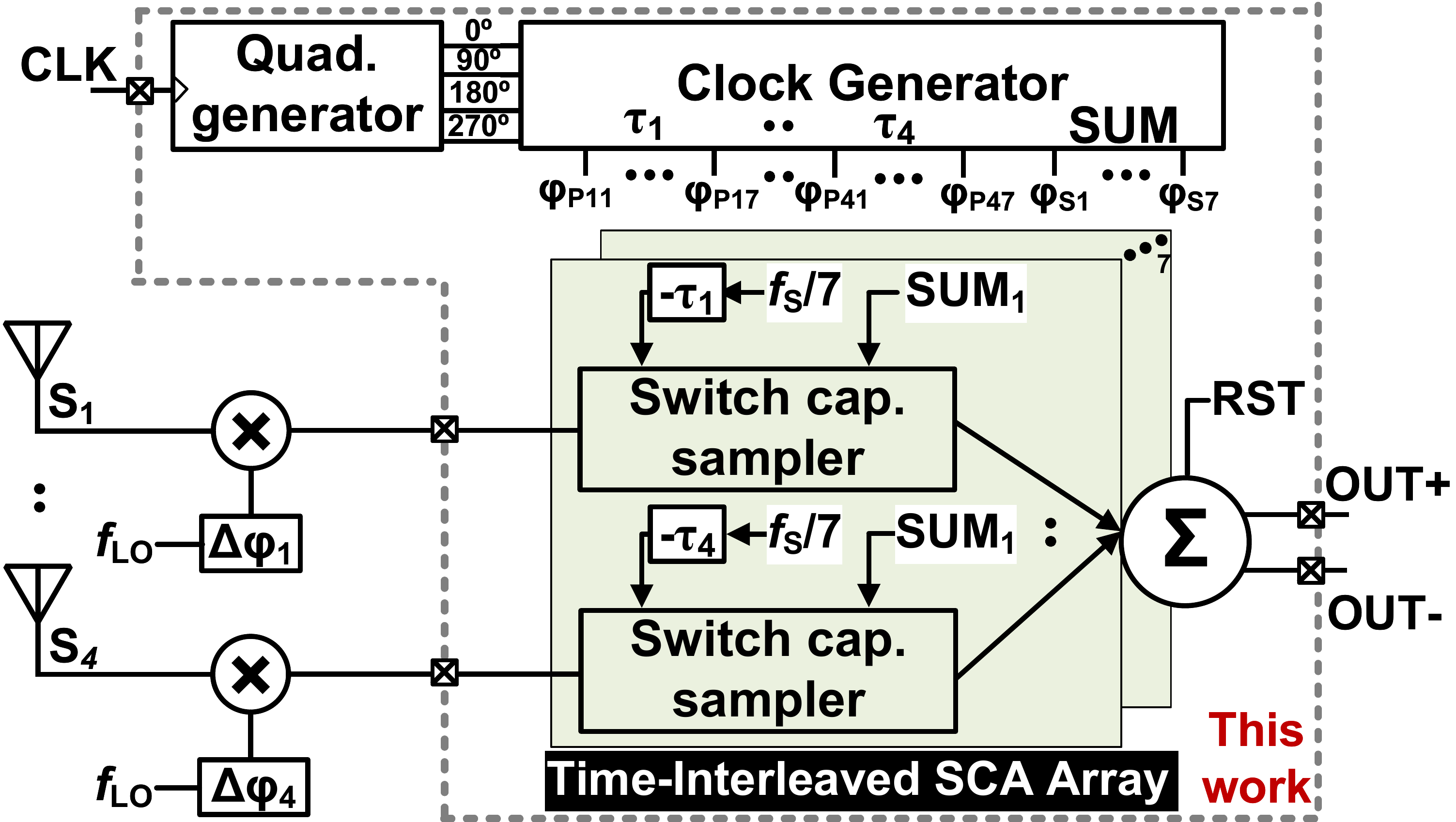}
    \vspace{-2mm}
    \caption{\small System architecture of proposed SSP.}
    \vspace{-4mm}
    \label{sys}
\end{figure}

%\begin{figure}[t]
%\centering
%         
%    \includegraphics[width=0.9\columnwidth]{Fig_ESSCIRC/delayconcept.pdf}
%    \vspace{-4mm}
%    \caption{\small Delay sampling compensation concept.}
%    \vspace{-4mm}
%    \label{concept}
%\end{figure}

\begin{figure}[t]
\centering
         
    \includegraphics[width=0.9\columnwidth]{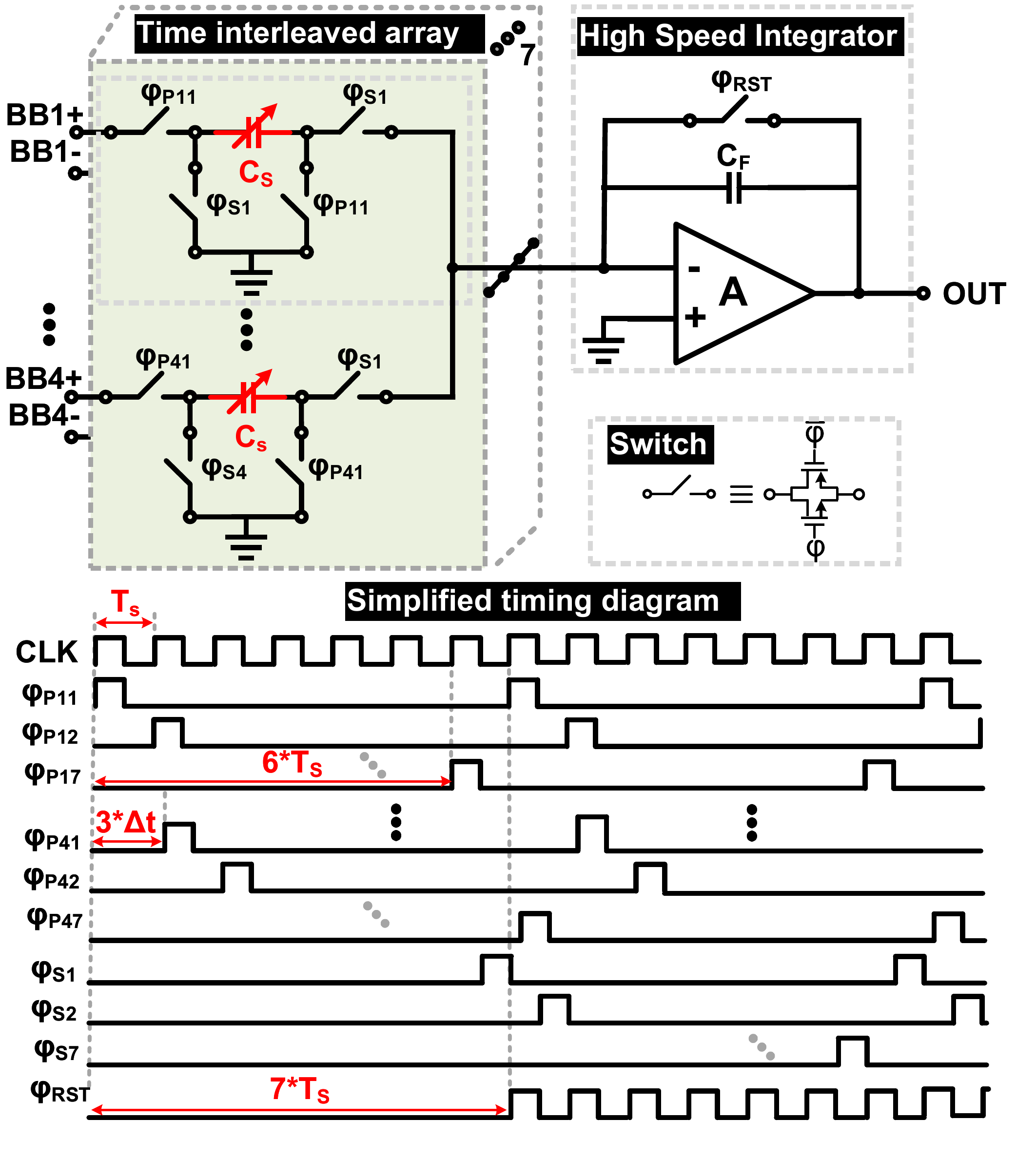}
    \vspace{-2mm}
    \caption{\small SCA array realization.}
    \vspace{-4mm}
    \label{SCA}
\end{figure}

\begin{figure}[t]
\centering
         
    \includegraphics[width=1\columnwidth]{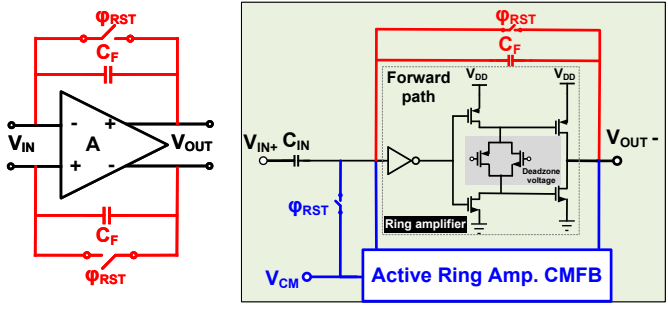}
    \vspace{-4mm}
    \caption{\small Ring amplifier (inset) embedded in SCA summer.}
    \vspace{-4mm}
    \label{ramp}
\end{figure}

\section{Proposed TTD SSP Architecture and Circuits }
This section describes the architecture and IC design of a TTD SSP that handles challenges described before.

The architecture for the proposed 4-element TTD SSP is shown in Fig.~\ref{SCA}. A time-interleaved ring-amplifier based switched-capacitor array (SCA) is used for signal combining. An interleaving factor of 7 was chosen to cover a large delay range while meeting the Nyquist rate sampling requirements of a wideband signal. The delay compensation  technique only requires digital logic, switches, and capacitors which can be well integrated in  CMOS technology  and benefit from process scaling. The SCA requires non-overlapping clocks for sampling, sum, and reset phases generated by a time-interleaver adapted from  \cite{ghaderi2019}. Each interleaved level has a conversion speed of roughly 228MHz (1.6GHz/7). The SCA uses 50fF sampling capacitor and optimally sized transmission gates for the switches in each path due to the common mode voltage being at roughly half the supply. Additionally, quadrature signals were assumed to allow for base band vector modulation, simplifying the design of the RF front end (vector modulator not included in this work for brevity).  

Summation at the OpAmp virtual ground limits the feedback factor affecting the maximum closed-loop bandwidth possible. While advanced techniques such as \cite{Naderi2018} can mitigate the aforementioned drawback, higher power, design complexity, and additional area limits its use. Ring amplifier (RAMP) is an emerging topology \cite{Hershberg2012} that aims to realize a low-voltage high-speed amplifier suited for advanced CMOS nodes. The RAMP provides high slew rate capabilities like that of underdamped systems initially, but settle to a desired value very quickly like that of critically damped systems. As shown in Fig.~\ref{ramp}, the RAMP was placed in a pseudo-differential configuration with its common mode feedback (CMFB) realized using another ring amplifier placed in the common mode (CM) feedback path with high deadzone voltage. This placement achieves a better common-mode error rejection ensuring the output node is faithfully represented for the entire summing operation \cite{Lagos2019}. 

%Vector modulation is enabled through the 5-bit binary switchable binary-weighted sampling capacitor for both in-phase and quadrature paths. The effective phase shift is approximated by the arctangent of the capacitor ratio of the Q-path to the I-path (i.e., $tan^{-1}$(C\textsubscript{Q}/C\textsubscript{I})). The switches are realized using transmission gate for higher swings. 
%Though the use of a single transistor switch (i.e., either NMOS or PMOS) saves the area and the layout effort, it could potentially result in the switch simply turning off for signals close to rails (i.e., supply and ground) degrading the performance significantly. 
%The bandwidth of the vector modulator is designed to be much larger than the desired signal determined by the switch on-resistance and the sampling capacitor. The simulated power consumption for a single path is only 125$\mu$W.    

\begin{figure}[t]
\centering
         
    \includegraphics[width=0.9\columnwidth]{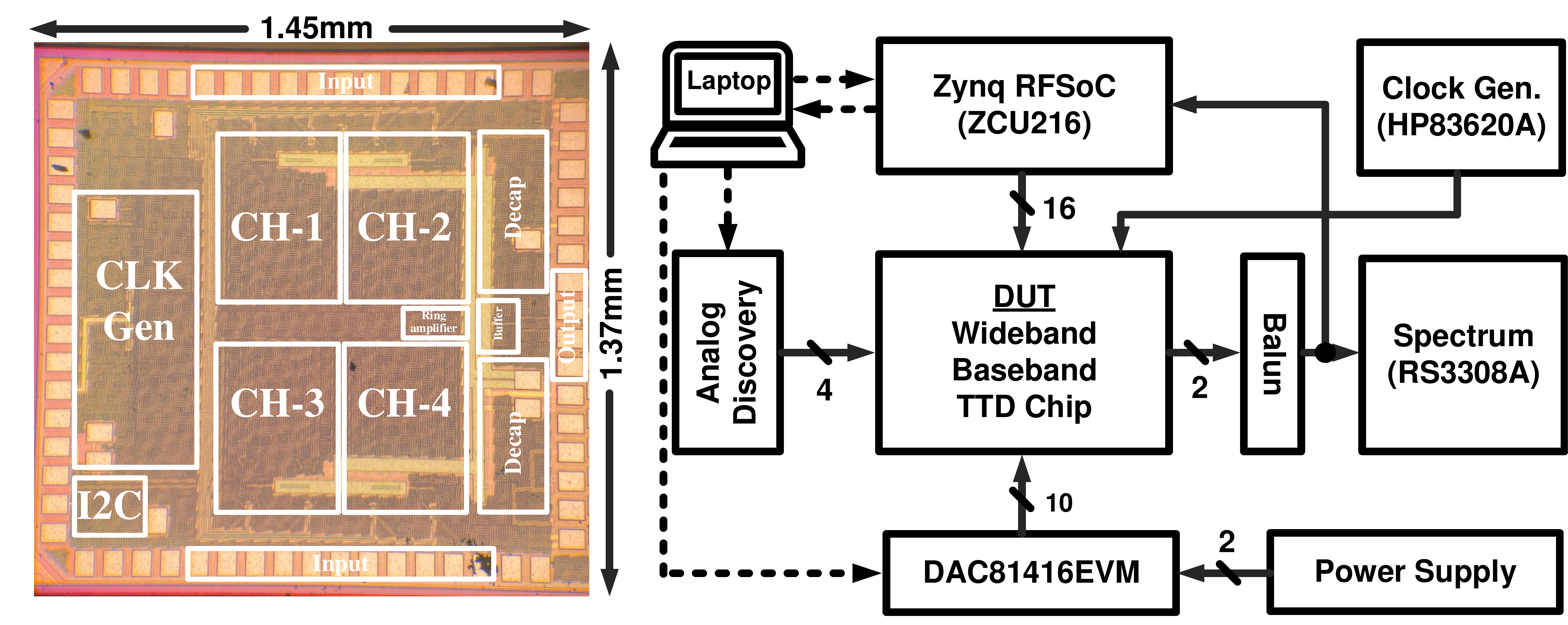}
    \vspace{-2mm}
    \caption{\small Die photo and simplified test setup.}
    \vspace{-4mm}
    \label{die}
\end{figure}

\begin{figure}[t]
\centering
    \includegraphics[width=1
    \columnwidth]{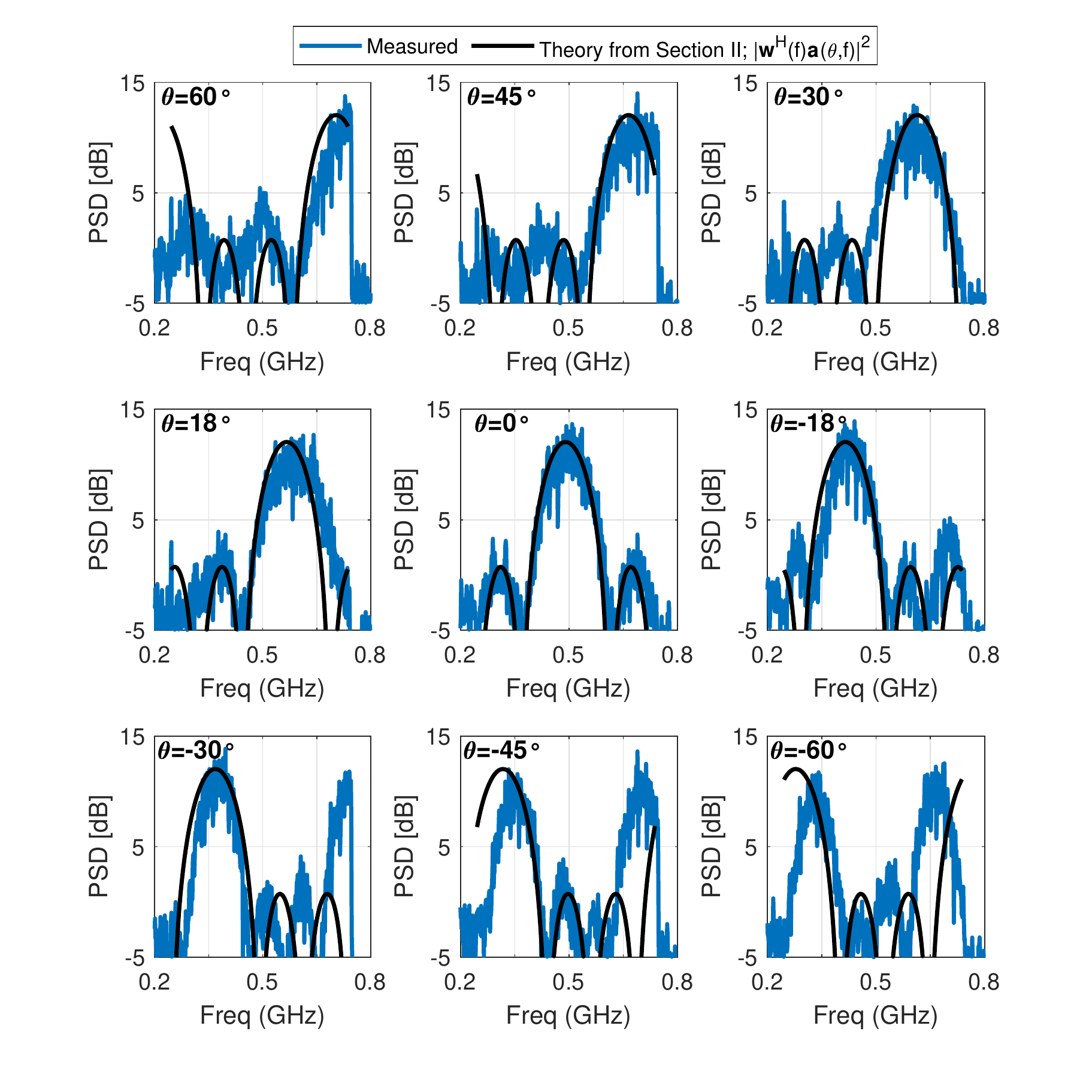} % Subhanshu changed from eps to pdf on 6_2_2021 to upload on arXiv
    \vspace{-12mm}
    \caption{\small Proof of fast beam training concept. With a fixed configuration of delay taps, the PSD of the received signal is uniquely determined by incidence angles $\theta$. Hence, $\theta$ can be inferred without time-consuming sequential beam sweep.}
    \vspace{-6mm}
    \label{BT}
\end{figure}

\begin{figure}[t]
    \begin{center}
        \includegraphics[width=0.4\textwidth]{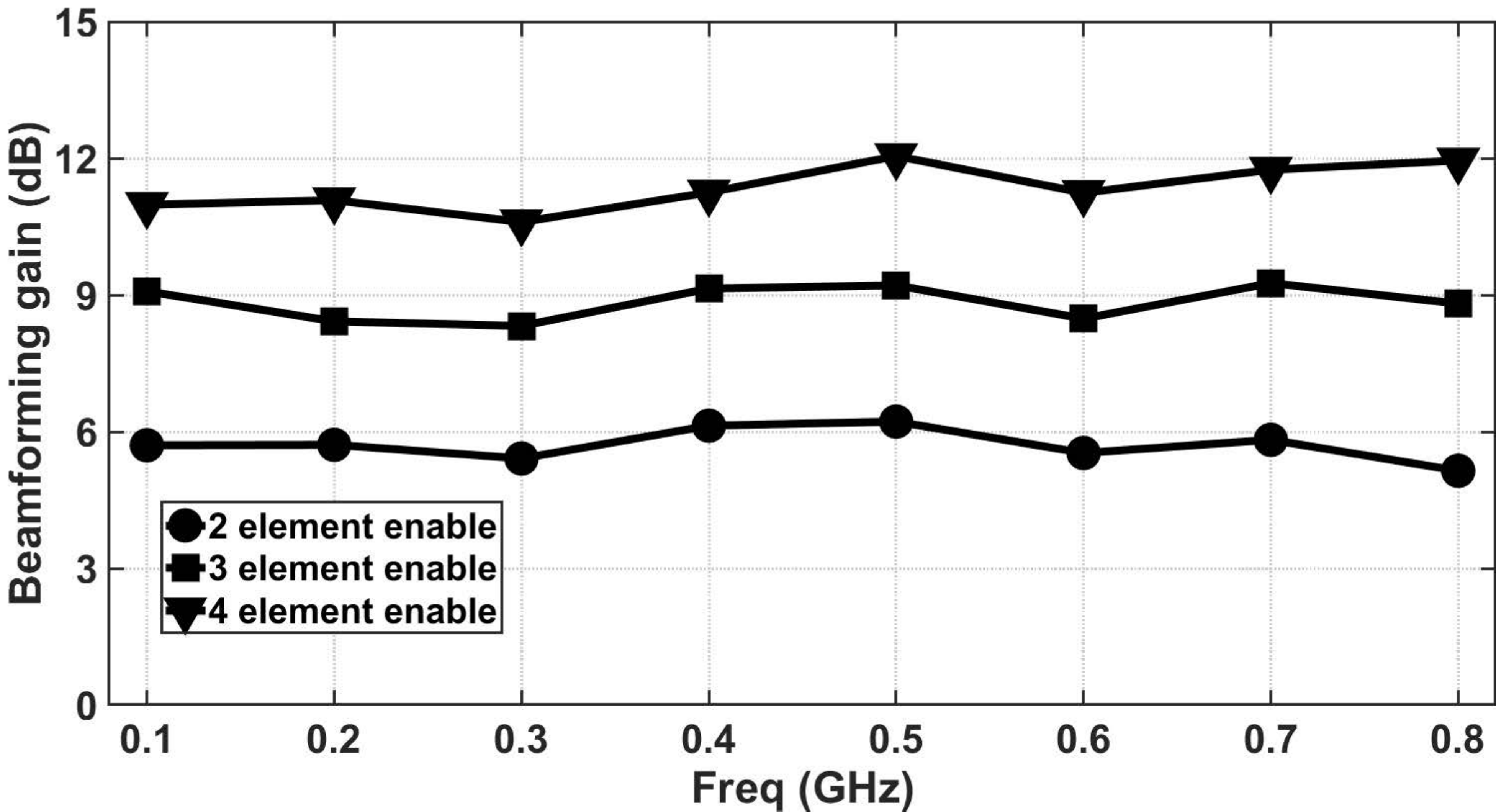}
    \end{center}
    \vspace{-5mm}
    \caption{Measured CW-tone beamforming gain.}
    \vspace{-4mm}
    \label{BFgain}
\end{figure}

%\subsection{Signal combining using hybrid amplifier}
%To combining the aligned BB signal, the integrator, which comprised with an operation amplifier and a feedback capacitor, is utilized. Classical amplifier topologies simply are not scalable due to the intrinsic loss of signal-to-noise ratio, and output voltage swing in advanced technology nodes caused by reduced supply voltage head room. 

\section{Measured Results} 
The proposed SSP was prototyped in 65nm CMOS occupying 1.98mm$^{2}$ area including  pads as shown in Fig.~\ref{die}. Fig.~\ref{die} also shows the test setup with the device-under-test (DUT). The Xilinx ZCU216 7.8GHz RF-DACs is used for both CW and modulated wideband waveform generation uploading MATLAB generated vectors. 

\begin{figure}[t]
    \begin{center}
        \includegraphics[width=0.4\textwidth]{Fig_ESSCIRC/beampattern_polar.pdf}
    \end{center}
    \vspace{-4mm}
    \caption{Measured beamforming pattern in polar format.}
    \vspace{-4mm}
    \label{BFpattern}
\end{figure}

%\begin{figure}[t]
%   \begin{center}
%       \includegraphics[width=0.24\textwidth]{Fig_ESSCIRC/VM.pdf}
%    \end{center}
%    \vspace{-4mm}
%    \caption{Measured response of 5-bit 4-quadrant vector modulator.}
%    \vspace{-4mm}
%    \label{VM}
%\end{figure}

\begin{figure}[t]
\centering
    \includegraphics[width=1\columnwidth]{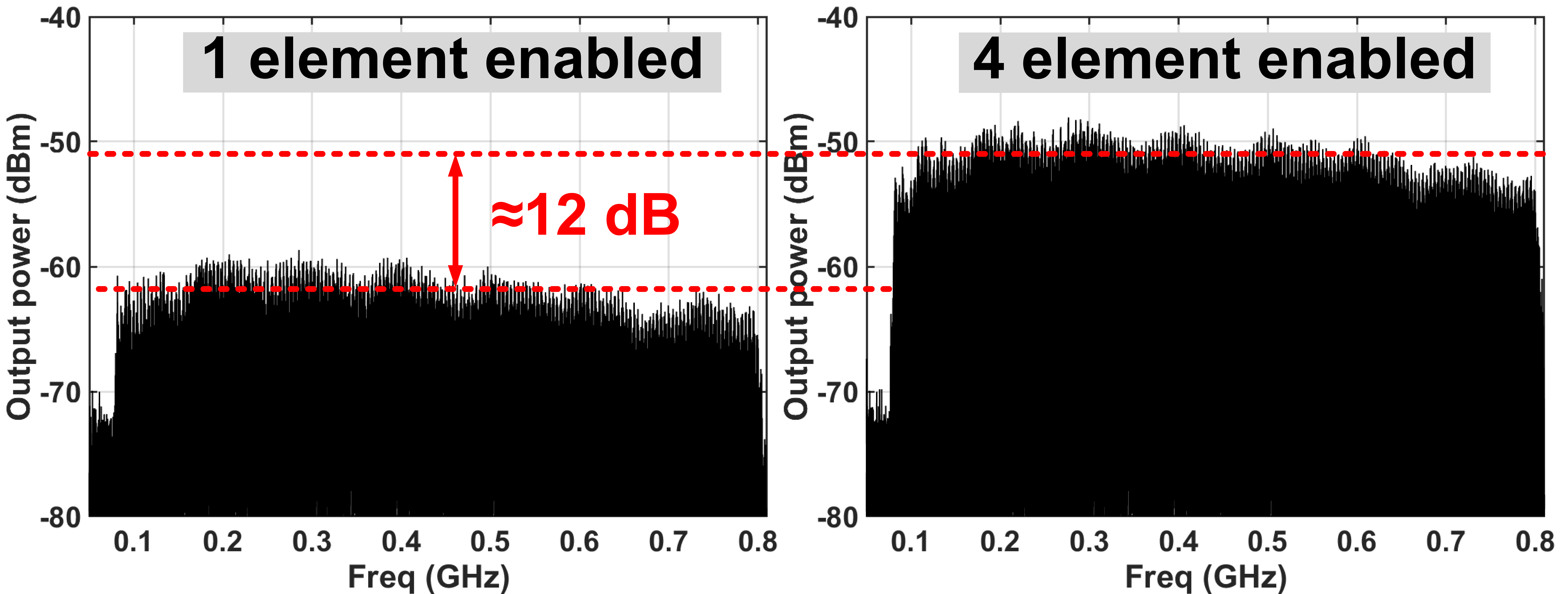}
    \vspace{-4mm}
    \caption{\small Measured wideband test with 720MHz span.}
    \vspace{-5mm}
    \label{WB8E6}
\end{figure}

\begin{figure}[t]
\centering
    \includegraphics[width=1\columnwidth]{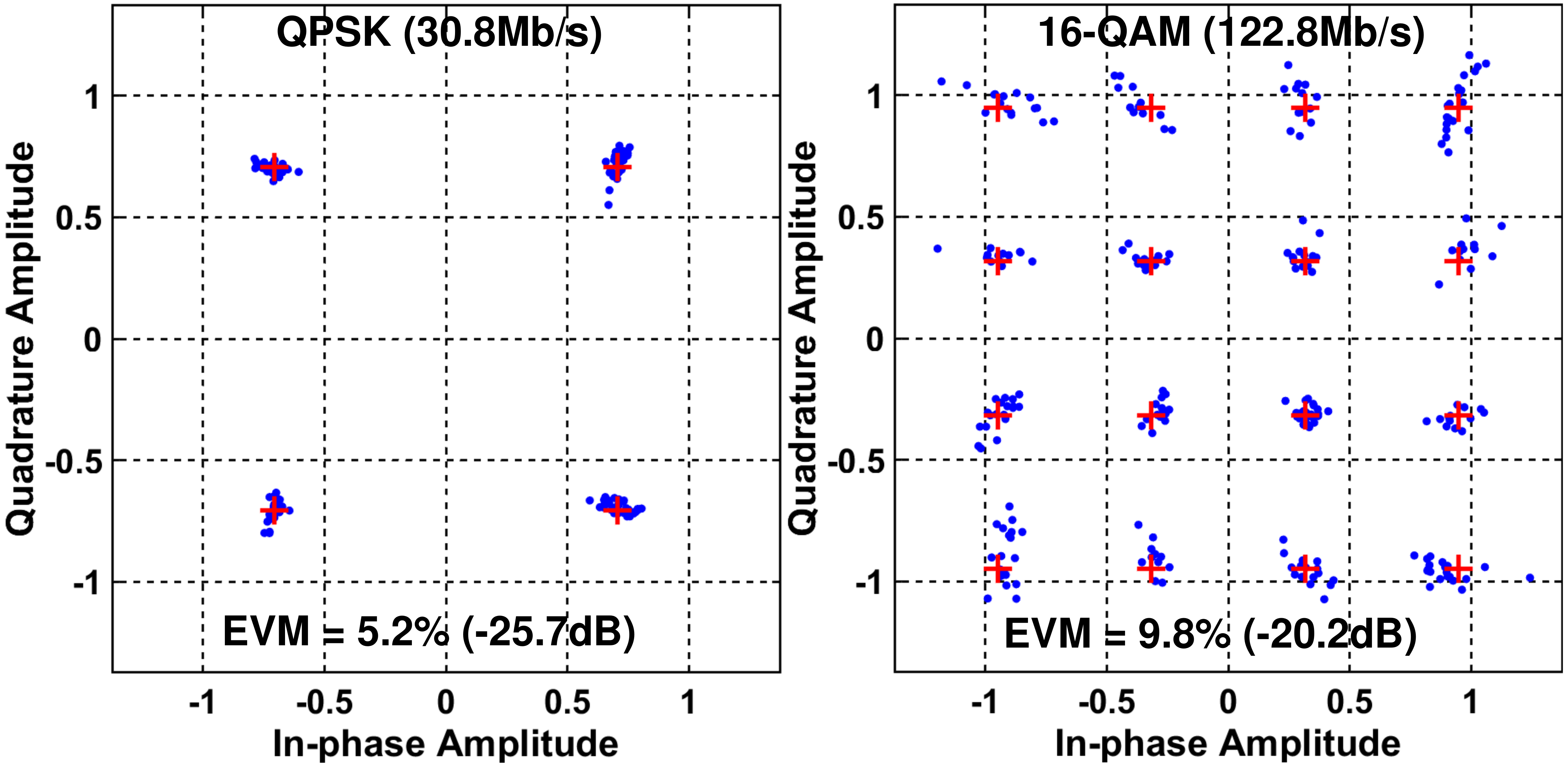}
    \vspace{-3mm}
    \caption{\small Measured EVM for QPSK and 16-QAM modulation.}
    \vspace{-5mm}
    \label{EVM}
\end{figure}

\textbf{Low-latency beam training} To prove the concept in Section~II, the input signals to the 4 channels of TTD IC are generated by RF-DAC which emulates the OFDM symbols received by a critically spaced linear array at $60$GHz with incident angle $\theta$, followed by down-conversion to intermediate frequency at $491.32$MHz. The signal has bandwidth $\text{BW} = 491.32$MHz and subcarrier spacing 960KHz. Each is loaded with a BPSK symbol. Although our beam training conceptually requires only one pilot OFDM symbol to measure the system frequency response and infer $\theta$, our experiment lacks the necessary synchronization. Hence, we intentionally let the same OFDM pilot symbol repeats itself and uses power spectrum density (PSD) estimation of the combined signal after DUT as indirect measurements of DUT frequency response. 
% in a wideband scenario with simplified DSP. Unlike the algorithm in \cite{Boljanovic2021}, where a cyclic-prefix based OFDM receiver uses a single wideband pilot to identify the optimal angle/beam based on the subcarrier with the maximum received power, our proof-of-concept aims to point out the difference in the received power spectral density (PSD) for different incidence angles when a repetitive sequence of OFDM symbols is used. 
The delay taps $\tau_n$ are set according to Section~II. Fig.~\ref{BT} presents the PSD to 9 different incidence angles $\theta$, in contrast with the theoretic frequency response of DUT to $\theta$ from Section~II. The results indicate the unknown incident angle $\theta$ can be identified by analyzing the frequency response of the array upon receiving a wideband pilot symbol, thanks to the TTD frequency-dependent antenna weight vector. The procedure accelerate beam training by avoiding the sequential switching-measuring as required by conventional method.
% Note that it contains a fixed delay tap configuration and there is no switching which is required in conventional beam training.
% The measured PSDs resemble the beam patterns due to the unique frequency-to-angle mapping, and the optimal angle can be found by identifying the frequencies with the maximum received power.

\textbf{Wideband communications}  Measured beamforming gain is plotted in Fig.~\ref{BFgain} for an incident angle of 0$^{\circ}$ with swept CW-tone from near DC to  800MHz, showing an uniform beamforming gain of $\sim6$dB, $\sim9$dB, and $\sim12$dB as the number of elements are increased from 2 to 4. For the angle of incidence testing, the beam patterns of the SSP are plotted in polar format (Fig.~\ref{BFpattern}), showing the effectiveness of the TTD SSP for $0^{\circ}, 30^{\circ}, 60^{\circ}$, and $-45^{\circ}$ . 
Fig.~\ref{WB8E6} shows a $\sim12$dB frequency-independent beamforming gain for a 720MHz wide bandwidth with four channels enabled demonstrating delay compensation with RAMP based signal combiner. The off-chip bandpass filter with 70MHz 3-dB high-pass corner attenuates the gain at lower frequencies. At higher frequencies, the observed gain roll-off is due to amplifiers in the ZCU216 RF-ADC.
Error vector magnitude (EVM) measurement with 30.8Mb/s QPSK signal and 122.8Mb/s 16-QAM signal are fed into the SSP for data communications performance evaluation. An EVM of 5.2\% and 9.8\% are obtained with four channels enabled as shown in Fig.~\ref{EVM}. In these results, the 5dB loss in the on-chip output buffer is not decoupled and that may limit the EVM performance. 

\begin{table}[t]
\vspace{0mm}
\centering
\caption{\small Comparison to prior-art ssp.  } 
\label{tab:FoMADC}
\vspace{-2mm}
{   
    \footnotesize { 
    \begin{tabular}{|p{0.55in}||p{0.32in}|p{0.32in}|p{0.32in}|p{0.32in}|p{0.32in}|} 
    \hline
        Parameters &  \cite{fikes2020} & \cite{ghaderi2019}  & \cite{Ghaderi2020ISSCC} & \cite{spoof2020} & This work  \\
        \hline
        \hline
        Method  & BB-TTD & BB-TTD & BB-TTD & RF-TTD &BB-TTD \\
        \hline
        Architecture   &  MIMO & MIMO & MISO & MISO  & MISO\\
        \hline 
        	\# of Elements  &  4 &  4 & 4 & 4 &  4 \\
        \hline
        Domain  &  Charge & Charge & Time & Charge  & Charge\\ 
        \hline 
        BW (MHz)  &  100 & 100 & 500 & 800 &  800\\ 
        \hline 
        Delay range (ns)  &  10 & 15 & 1 & 5 &  3.8\\ 
        \hline 
          Digital intensive  &  Yes & No & Yes & Yes & Yes\\ 
        \hline 
          Beamtraining  &  No & No & No & No &  Yes \\ 
        \hline 
          Power(mW)  &  $70^*$ & 52 & $40^\dagger$ & 70 & $29^\S$ \\ 
        \hline 
          Area ($\text{mm}^2$)  &  $0.5^*$ & 0.82 & 0.9 & 1.2  & $1.98^\sharp$\\
        \hline 
          Technology (nm)  &  65 & 65 & 65 & 28  & 65\\ 
        \hline 
    \end{tabular} \\
}
}
\vspace{-1mm}
\flushleft
*Baseband delay, mixers, and LO; $^\dagger$incl. ADC power; $^\S$excl. test buffers; $^\sharp$incl. vector modulator
\vspace{-3mm}
\end{table}

Table~\ref{tab:FoMADC} summarizes the critical parameters for the proposed TTD SSP and compares with state-of-the-art. Though \cite{spoof2020} demonstrated similar bandwidth as the proposed work, the use of a RF sampling mixer and digitally-controlled delay line will limit its application at mmWave frequencies. In contrast, the proposed work has the ability to connect with different RF and mmWave front end downconverters relaxing the overall system design complexity for TTD arrays.  

\section{Conclusions} 
This work demonstrates for the first time the fast beam-training algorithm for TTD arrays leveraging frequency-dependent search beams to sound all directions simultaneously which greatly reduces beam training latency. In addition, the proposed architecture supports  wideband data communications with large delay-bandwidth product using fast slewing wideband ring amplifier for efficient signal combining  in the baseband switched-capacitor array. A 3.8ns delay compensation across 800MHz bandwidth is demonstrated with EVM of $<10\%$ supporting 16-QAM. Prototyped in 65nm CMOS with 1.98mm$^2$, the TTD SSP consumes 29mW only.

\bibliographystyle{IEEEtran}
\bibliography{ref}

\end{document}